\documentclass{class}

\usepackage{times}
\usepackage{hyperref}
\usepackage[compact]{titlesec}
\usepackage{tikz}
\usepackage{amsmath}
\usepackage{xspace}
\usepackage{graphicx}
\usepackage{algorithm} 
\usepackage{algpseudocode} 
\usepackage[normalsize,TABBOTCAP]{subfigure}
\usepackage{multirow}
\usepackage{arydshln}
\usepackage{booktabs}
\usepackage{kotex}
\usepackage{colortbl}
\usepackage{upgreek}        
\usepackage{unitsdef}
\usepackage{nopageno}
\usepackage{makecell}
\usepackage{tabularx}
\usepackage{pifont}
\usepackage{tikz}
\usepackage{threeparttable}
\usepackage{tablefootnote}
\usepackage{caption}
\usepackage{enumitem}

\setitemize{nosep,leftmargin=*}
\hypersetup{pdfstartview=FitH,pdfpagelayout=SinglePage}

\newcommand{\sysname}{Lovelock\xspace}
\newcommand{\parab}[1]{\vspace*{0.25ex}\noindent\textbf{#1}}

\newtoggle{showmarks}
\togglefalse{showmarks}
\iftoggle{showmarks}{
\newcommand{\seojin}[1]{[\textcolor{orange}{\sf\textit{#1 - Seojin}}]}
\newcommand{\gwkim}[1]{[\textcolor{brown}{\sf\textit{#1 - Geonwoo}}]}
\newcommand{\ramesh}[1]{[\textcolor{red}{\sf\textit{#1 - Ramesh}}]}
\newcommand{\kshen}[1]{[\textcolor{purple}{\sf\textit{#1 - Kai}}]}
\newcommand{\dculler}[1]{[\textcolor{purple}{\sf\textit{#1 - David}}]}
\newcommand{\fatma}[1]{[\textcolor{green}{\sf\textit{#1 - Fatma}}]}
\newcommand\red[1]{\textcolor{red}{#1}}
\newcommand\redstrike[1]{\red{\sout{#1}}}
\newcommand\green[1]{\textcolor{\green}{#1}}
\newcommand\greenstrike[1]{\green{\sout{#1}}}
\newcommand\orange[1]{\textcolor{orange}{#1}}
\newcommand\orangestrike[1]{\orange{\sout{#1}}}
\newcommand\blue[1]{\textcolor{blue}{#1}}
\newcommand\bluestrike[1]{\blue{\sout{#1}}}
\newcommand\purple[1]{\textcolor{purple}{#1}}
\newcommand\purplestrike[1]{\purple{\sout{#1}}}
}{
\newcommand{\seojin}[1]{}
\newcommand{\gwkim}[1]{}
\newcommand{\ramesh}[1]{}
\newcommand{\kshen}[1]{}
\newcommand{\dculler}[1]{}
\newcommand{\fatma}[1]{}
\newcommand\red[1]{#1}
\newcommand\redstrike[1]{\unskip}
\newcommand\green[1]{#1}
\newcommand\greenstrike[1]{\unskip}
\newcommand\orange[1]{#1}
\newcommand\orangestrike[1]{\unskip}
\newcommand\blue[1]{#1}
\newcommand\bluestrike[1]{\unskip}
\newcommand\purple[1]{\unskip}
\newcommand\purplestrike[1]{\unskip}
}

\definecolor{Mycolor2}{HTML}{009933}

\begin{document}

\title{\sysname: Towards Smart NIC-hosted Clusters}

\author{
Seo Jin Park{$^{12}$}\quad Ramesh Govindan {$^{12}$}\quad Kai Shen$^1$\quad David Culler{$^1$} \\ Fatma Özcan{$^1$}\quad Geon-Woo Kim{$^3$}\quad Hank Levy{$^1$}\\\\
$^1$ Google \quad $^2$ University of Southern California \quad $^3$ UT Austin
}

\maketitle

\begin{abstract}

Traditional cluster designs were originally server-centric, and have evolved recently to support hardware acceleration and storage disaggregation. In applications that leverage acceleration, the server CPU performs the role of orchestrating computation and data movement and data-intensive applications stress the memory bandwidth.
Applications that leverage disaggregation can be adversely affected by the increased PCIe and network bandwidth resulting from disaggregation. In this paper, we advocate for a specialized cluster design for important data intensive applications, such as analytics, query processing and ML training. This design, \sysname, replaces each server in a cluster with one or more headless smart NICs. Because smart NICs are significantly cheaper than servers on bandwidth, the resulting cluster can run these applications without adversely impacting performance, while obtaining cost and energy savings.

\end{abstract}

\section{Introduction}

Until recently, datacenter clusters were server-centric: servers with significant compute and storage, connected by a high-speed fabric, enabled massively parallel data processing applications. In these, a single application instance can recruit tens of thousands of worker nodes to load and process input data in parallel, followed by shuffling results through the network fabric. For large datasets, such applications can consume massive computational power and bandwidth.

More recently, cluster designs have evolved to accommodate \textit{acceleration} and \textit{disaggregation}. Custom hardware can be more efficient for some workloads (e.g., ML training and inference, video encoding/decoding), so cluster designs now include accelerators attached to servers. Clusters also disaggregate storage~---~dedicated servers for serving storage requests over the network~---~in an effort to scale storage and compute independently. Recent research suggests that future clusters will disaggregate memory~\cite{aifm, carbink, legoos} and accelerator access~\cite{audibert2022case, cachew} as well to circumvent the problem of right-shaping resources to tasks.

In this paper, we take the position that with the advent of acceleration and disaggregation, for several important applications (analytics, query processing, ML training), a server-centric design may no longer be necessary (\S\ref{sec:background}). In servers with increasing core counts, cores contend for network, memory and PCIe bandwidth. This is exacerbated by disaggregation, which increases traffic on the PCIe bus and the network. For applications that use accelerators heavily, the server CPU is reduced to the role of a coordinator, merely orchestrating computations and data movement to avoid accelerator stalls. 


\begin{figure}[t]
\centering
\includegraphics[width=\linewidth]{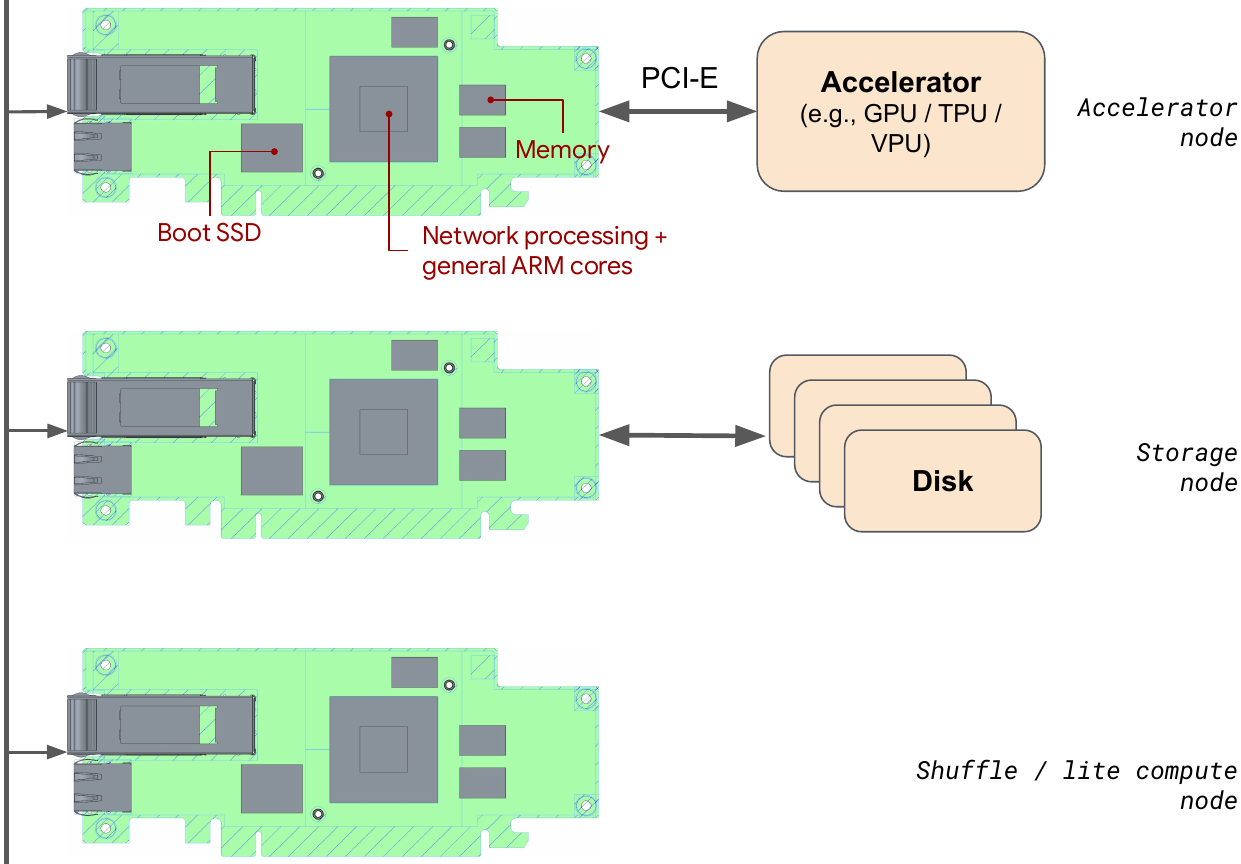}
\caption{Architecture of \sysname}
\label{fig:arch}
\end{figure}

Instead, we argue that it is more cost-effective and energy-efficient to design specialized server-less clusters for these applications. 
Our proposed cluster design, \sysname, replaces servers with headless smart NICs (Figure~\ref{fig:arch}). Today's smart NICs (e.g., Intel IPU E2000~\cite{e2000}, Bluefield DPU~\cite{DpuWhitePaper}, AMD Pensando~\cite{pensando}) 
were originally designed to offload networking and infrastructure tasks, but they possess enough compute (e.g., 16 ARM cores), memory (16-48 GBs) and PCIe connectivity to serve as a platform for disaggregation (\S\ref{sec:background}). A smart NIC also costs substantially less in capital and operating (energy) expenditures ---
e.g., \$1500 vs.\ \$10500 (7x) and 65W vs.\ 728W (11x) respectively in \cite{DpuWhitePaper}. Thus, even if a \sysname cluster were to replace each server with \textit{multiple} smart NICs, it could still be substantially cheaper and more energy-efficient than a server-centric design.

\begin{table}[]
\scriptsize
\centering

\resizebox{\columnwidth}{!}{%
\begin{threeparttable}
\begin{tabular}{|l|r|r|c|r|r|}
\hline
 \multicolumn{1}{|c|}{ } &
 \multicolumn{1}{c|}{\textbf{\makecell{Cores\\ vCPUs}}} &
 \multicolumn{1}{c|}{\textbf{NIC}} &
 \multicolumn{1}{c|}{\textbf{DRAM}} &
 \multicolumn{1}{c|}{\textbf{\makecell{NIC bw \\ per core}}} &
 \multicolumn{1}{c|}{\textbf{\makecell{DRAM bw \\ per core}}}
 \\ \hline
 \multicolumn{6}{|c|}{\textbf{Cloud host systems}} \\ \hline
 \makecell{Google Cloud N1 \\ 2x Intel Skylake} & 96 & 100Gbps & \makecell{2x 6-ch \\ DDR4} & 0.13\,GB/s & 2.67\,GB/s \\ \hline
 \makecell{Google Cloud N2d \\ 2x AMD Milan} & 224 & 100Gbps & \makecell{2x 8-ch \\ DDR4} & 0.06\,GB/s & 1.83\,GB/s \\ \hline
 \makecell{AWS M6in \\ 2x Intel Ice Lake} & 128 & 200Gbps & \makecell{2x 8-ch \\ DDR4} & 0.20\,GB/s & 3.20\,GB/s \\ \hline
 \makecell{Google Cloud C3 \\ 2x Sapphire Rapids} & 176 & 200Gbps & \makecell{2x 8-ch \\ DDR5} & 0.14\,GB/s & 3.49\,GB/s \\ \hline
 \makecell{AMD Genoa\tnote{1} \\(1x EPYC 9654)} & 192 & 200Gbps & \makecell{12-ch \\ DDR5} & 0.13\,GB/s & 2.40\,GB/s \\ \hline
 \multicolumn{6}{|c|}{\textbf{Smart NICs}} \\ \hline
 IPU E2000~\cite{intel-e2000-isscc-2023} & 16 & 200Gbps & \makecell{3-ch \\ LPDDR4} & 1.56\,GB/s & 6.40\,GB/s \\ \hline
 Bluefield v3~\cite{bluefield-3-datasheet} & 16 & 400Gbps & \makecell{2-ch \\ DDR5} & 3.13\,GB/s & 5.60\,GB/s\\ \hline
\end{tabular}
\begin{tablenotes}
\item[1] ``AMD Genoa'' is not yet released on public clouds, so we assumed 1 socket of EPYC paired with a 200Gbps NIC and the highest possible memory bandwidth.
\end{tablenotes}
\end{threeparttable}
}
\caption{Network and DRAM bandwidths per core of different platforms.  The reported bandwidths are theoretical, not effective bandwidths by measurements.  The theoretical DDR bandwidths were computed using DDR transfer rates if reported publicly, or the max transfer rate of the respective DDR technology otherwise.}
\label{tbl:bandwidthPerCompute}
\end{table}


\sysname can improve efficiency without compromising the performance of data-intensive applications
because smart NICs offer substantially higher network and memory band\-width-to-compute ratios than traditional servers (Table~\ref{tbl:bandwidthPerCompute}). The high network bandwidth enables faster network transfers for applications that leverage disaggregation, compensating for the lower CPU speeds on the smart NIC (\S\ref{sec:results}). The higher memory bandwidth allows each core of a smart NIC to be more efficient, relative to a server core. Using a simple model of cost and power (\S\ref{sec:model}), we show that for certain applications,
\sysname can reduce capital cost by 21\%-71\% and energy use by 23\%-80\%.

These preliminary, back-of-the-envelope analyses are encouraging, but require significant work in improving the design of smart NICs, increasing the efficiency of the network stack and isolation mechanisms, and scaling disaggregated applications efficiently (\S\ref{sec:discuss}).





\section{Background and Motivation}
\label{sec:background}

We begin with a brief background on smart NICs, then make several observations that motivate our work.

\subsection{Smart NICs}
\label{sec:smartnic}

Originally, smart NICs were designed to offload packet processing from the host CPU with the goal of preserving CPU cycles for application workloads. Early smart NICs, for example, supported TCP segmentation and re-assembly, measurement, and access control, and could also be programmed to perform general packet match-action tasks~\cite{netronome, liquidio}. Since then, smart NICs have evolved to have on-board general-purpose compute and significant memory, to the point where they are considered generalized data processing units (DPUs) or infrastructure processing units (IPUs). 

\begin{figure}[t]
\centering
\includegraphics[width=0.7\linewidth]{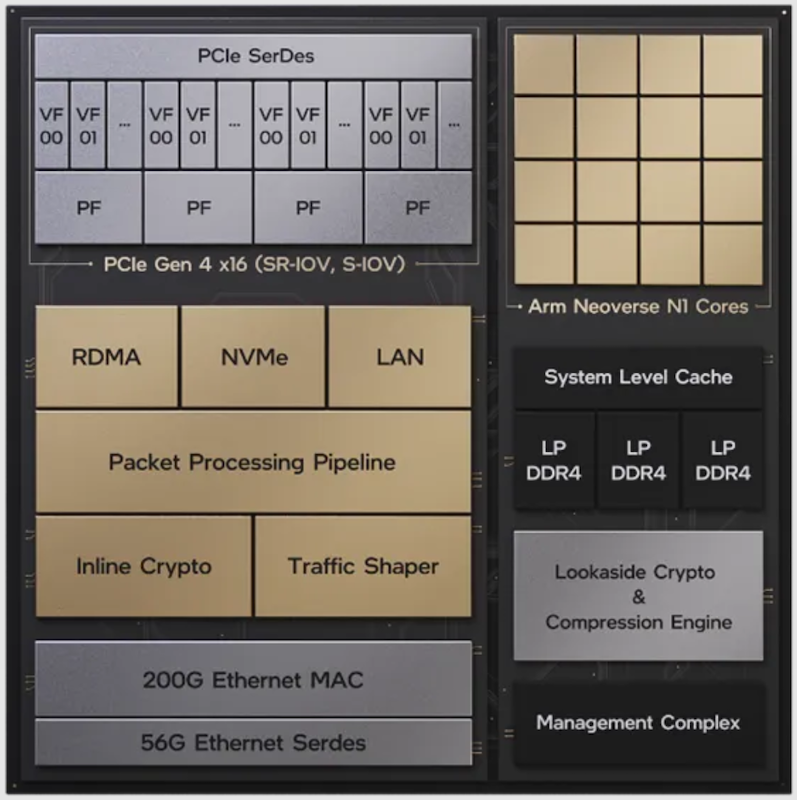}
\caption{Intel IPU E2000 design \protect\cite{e2000}}
\label{fig:intelIPU}
\end{figure}

Figure~\ref{fig:intelIPU} shows the components of the latest smart NIC from Intel, IPU E2000. It has a built-in processor with 16 Arm cores and a low-power DRAM (LP-DDR4). On this hardware, today's smart NICs run a commodity operating system (such as Linux), and can support power-efficient execution of general purpose computations without requiring significant code modifications. Beyond these, smart NICs have specialized hardware for common tasks. For example, the E2000 has a programmable match-action packet processing pipeline to implement access control, NAT, or congestion control in hardware. 
Smart NICs commonly support acceleration for encryption and compression, two operations that consume significant CPU cycles in datacenters~\cite{hyperscaleBigDataProcessing}. These accelerators free up the use of compute cores for other tasks, a capability we exploit in this paper. 
It is also common for smart NICs to provide PCIe connectivity to attach accelerators, storage, and other peripherals.

\subsection{Motivation}
\label{sec:motivation}

Several trends in datacenter computing and data-intensive applications motivate our work.


\parab{Increasing core counts create bandwidth bottlenecks.} Cl\-oud operators sell CPU cores (accompanied with 4\,GB or so DR\-AM per core) to customers. To reduce per-core capital cost, it is now common for a host system to have hundreds of cores. Consequently, the system network and memory bandwidths are now shared with more cores, which have increasingly bottleneck-ed application performance.


\parab{Weak isolation and its impact on tail latency.} To utilize hundreds of cores, a host now has to serve multiple independent applications (or cloud VM instances)~\cite{wsm, borg}. 
Applications are typically assigned  dedicated CPU cores and some reserved memory capacity. However, other resources, such as memory bandwidth, last level cache, PCIe bandwidth, and network bandwidth, are still shared. Contention on those sha\-red bandwidth resources can degrade application performance. This can potentially be alleviated using class-of-service or QoS enhancements to some of these resources (e.g., ToS for network traffic, and isolation mechanisms for other shared resources~\cite{caladan,heracles,parties}), but in practice, these provide weak isolation. Even mild contention can result in higher tail latency and worse end-to-end performance especially for data-intensive workloads targeted in this paper.

\noindent\textbf{Disaggregation increases PCIe and network traffic.} Disaggregating memory and storage help independently scale storage and computation, and can increase memory and storage efficiency. However, disaggregation can add significant traffic to the network and the host-to-NIC PCIe bus. On a disaggregated host, memory traffic must traverse the PCIe bus and the network. Similarly, disaggregated storage traffic consumes additional network bandwidth, and additional PCIe bandwidth at the remote end. Increasingly, the PCIe bus is becoming a significant bottleneck for applications~\cite{pcie4net1, pcie4net2}. This is exacerbated by the increase in host-attached accelerators for graphics, video, and machine learning.


\parab{The changing role of the host CPU.} With increased use of hardware acceleration, the role of the CPU on a server in a data center has been changing. Increasingly, the CPU runs application logic that rarely performs intensive computation but focuses on coordinating computation on the accelerators and on transferring data between these devices and disaggregated memory and storage to avoid stalls on accelerators.








\section{\sysname: Clusters for Data-intensive Workloads}
\label{s:arch}

Motivated by these trends, we explore a novel architecture, \sysname, for a specialized pod or cluster for some data-intensive workloads. A \sysname cluster is distinguished by the complete absence of server-class machines (Figure~\ref{fig:arch}). Instead, in \sysname, smart NICs perform the functions of servers in a traditional cluster. Thus, the cluster consists entirely of network-attached smart NICs. 

In addition, each smart NIC may have one or more additional peripherals connected over PCIe, such as accelerators and SSDs. Specifically, we envision each node in a \sysname cluster to be one of: an \textit{accelerator node} which contains an attached GPU, TPU, video processor, crypto accelerator, etc; a \textit{storage node} that contains several physical storage devices (e.g., SSDs or HDDs) and serves storage requests over the network; or a \textit{lite compute} node without peripherals used entirely for lightweight computations or data shuffles.


\sysname is a specialized architecture for a specific subset of applications (bandwidth, not compute, bound applications) that leverages the potential cost and power benefits that smart NICs provide. It leverages the trends described in Section~\ref{sec:motivation} as follows:
\begin{itemize}
    \item Per-core memory bandwidth and shared cache are larger in \sysname, resulting in higher per-core performance relative to cores on traditional servers.
    \item Each smart NIC now serves fewer (or a single) applications, lessening the chance of contending on shared network/memory/PCIe bandwidths.
    \item \sysname improves disaggregation by having higher network bandwidth and removing the PCIe traffic between NIC and host CPUs. \blue{For example, the IPU E2000 uses a special mesh fabric, instead of PCIe, between the network processor and its ARM cores.)}
    \item For applications in which the CPU simply acts as a coordinator, the minimal compute on DPUs in \sysname is a better fit in terms of power consumption.
\end{itemize}

\vspace{0.5ex}
Because a smart NIC can be an order of magnitude cheaper and more power efficient than a traditional server, a \sysname cluster can \textit{scale out} smart NICs~---~replace one server with multiple smart NICs~---~to achieve comparable application performance while still being more cost and energy efficient (\S\ref{sec:model}). This scale out results in a cluster with higher aggregate bandwidth, which can benefit some applications (\S\ref{sec:results}).

In this paper, we take a first step towards understanding the feasibility of \sysname. Specifically, we:
\begin{itemize}
    \item Explore, using very simple analytic models, the cost and power-efficiency gains from \sysname relative to traditional clusters with servers (\S\ref{sec:model})
    \item Describe, and substantiate with measurements, a few applications that can benefit from \sysname (\S\ref{sec:results}).
    \item Discuss directions for future research  (\S\ref{sec:discuss}).
\end{itemize}

\section{Energy and Cost Modeling}
\label{sec:model}

\begin{figure*}
\centering
\includegraphics[width=\linewidth]{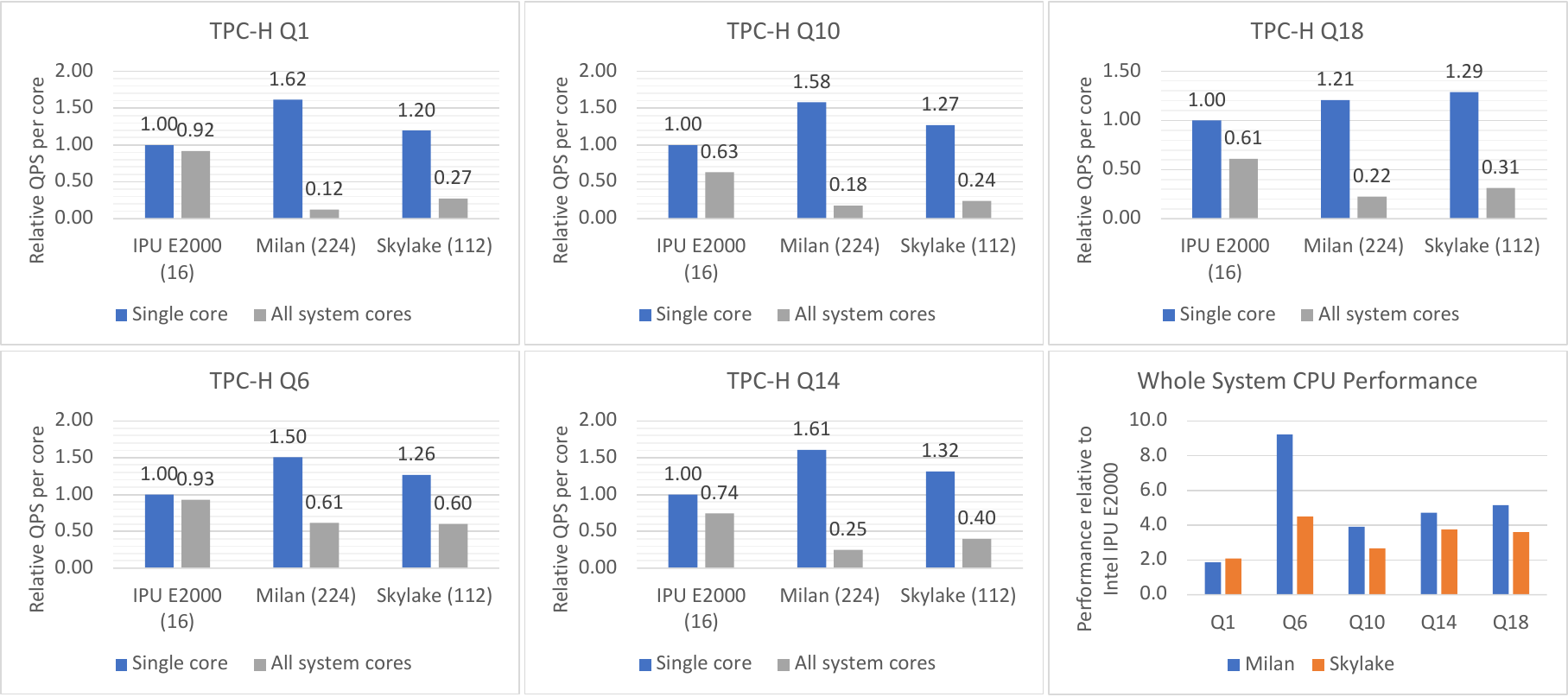}
\caption{Per-core performance when each core (SMT) independently executes a TPC-H query (so, no synchronization among cores). A proprietary analytics execution engine and TPC-H scale factor 1 (about 1 GB of data when uncompressed) were used. Performance was measured by execution time and normalized by the performance of Intel IPU E2000 when used only one core.}
\label{fig:voxel-cpu-eff}
\end{figure*}

The cost and energy benefits of \sysname are somewhat difficult to quantify, in part due
to the scarcity of public information on capital costs,
and because both cost and power advantages can change over time. We use an analytical model to get a preliminary understanding of \sysname's benefits.
Our analysis is best-effort given the available public information.

\noindent\textbf{Notation.} Suppose $c_s$ is the capital cost of a server relative to that of a SmartNIC and $p_s$ the power draw of a server relative to a SmartNIC. Analogously, let $c_p$ be the cost of PCIe devices (again, relative to the SmartNIC) attached to a server in a traditional cluster, or to the SmartNIC in \sysname, and let $p_p$ be their relative power. Now, a \sysname cluster is likely to be slower than a traditional cluster, which presents cluster designers with two degrees of freedom: they can provision $\phi$ times more SmartNICs 
than a traditional cluster servers and/or accept a slow-down $\mu$ on application execution. These two terms are knobs that designers can use to trade-off cost, power, and application performance\footnote{\blue{For ease of exposition, we have omitted fabric costs from the model. However, the model can be extended easily to account for increased fabric costs; we discuss this in \S\ref{sec:network-bandwidth} and \S\ref{sec:discuss}.}}.

\noindent\textbf{Cost and energy saving.} Using the notation above, we can approximate the ratio of the capital cost of a traditional cluster to the cost of a \sysname cluster as:
\begin{equation} \label{eq:cost}
 \frac{c_s + c_p}{\phi + c_p}
\end{equation}

and the ratio of the power draw of a traditional cluster to that of a \sysname cluster as:
\begin{equation} \label{eq:energy}
\frac{p_s + p_p}{\mu(\phi + p_p)}
\end{equation}

A recent white paper from NVIDIA on their Bluefield v2 SmartNIC~\cite{DpuWhitePaper} suggests $c_s\approxeq 7$ and $p_s \approxeq 11$. A \sysname cluster without PCIe devices that runs bandwidth-intensive applications and has 3$\times$ as many SmartNICs as servers (i.e., $\phi=3$) and runs these applications 20\% slower (i.e., $\mu=1.2$) is still 2.3$\times$ cheaper and uses 3.1$\times$ less energy!

For a cluster with PCIe devices, assume that the cost and power of PCIe devices is about 75\% of the total system\footnote{Rough estimate based on commercial systems with 4 GPUs/server.}. Then, using $c_s = 7$, $p_s=11.2$ in \cite{DpuWhitePaper} again, the cost and power ratios for PCIe devices will be $c_p = 7 \times \frac{0.75}{1 - 0.75} = 21$ and $p_p = 11.2 \times \frac{0.75}{1 - 0.75} = 33.6$. A \sysname cluster with 1 smart NIC in place of 1 server (i.e., $\phi=1$) and without any slowdown, has a 1.27x cost saving and 1.3x energy reduction. If \sysname is configured to use 2x more smart NICs ($\phi=2$) to improve application performance by 10\% ($\mu=0.9$), it can save 1.22x on cost and 1.4x on energy.

In \S\ref{sec:results}, we use this model to quantify benefits of \sysname.

\section{Initial Study Results}
\label{sec:results}

In this section, we explore the following hypotheses with respect to  \sysname clusters: 
\begin{itemize}
    \item Smart NIC CPU cores can outperform traditional hosts for memory-bandwidth-intensive workloads (\S\ref{sec:cpu-efficiency}).
    \item Higher network bandwidth can improve query processing performance at lower cost (\S\ref{sec:network-bandwidth}).
    \item They have CPU and memory capacity to drive high performance accelerators such as GPUs/TPUs, and giving higher bandwidth per accelerator reduces accelerator stalls (\S\ref{sec:accel-feasibility}).
\end{itemize}

\subsection{Higher CPU Core Efficiency}
\label{sec:cpu-efficiency}

Smart NICs have 7-11x fewer cores than traditional systems (Table~\ref{tbl:bandwidthPerCompute}). If a smart NIC is $\sim$7x cheaper than a traditional host~\cite{DpuWhitePaper}, a \sysname cluster with compute capacity comparable to a traditional cluster will have no cost advantages.

However, we anticipate that, at least for data-intensive workloads, each core of smart NIC can outperform a traditional host's core because it has higher memory bandwidth and larger L3 cache. To quantify this, we run TPC-H benchmarks with scale factor of 1 on an analytics execution engine to show that contention on shared bandwidth impacts traditional host core performance much more than a Smart NIC core.

We use three different systems for this evaluation. \textit{IPU E2000} has 16 ARM N1 cores and 48 GBs of memory. \textit{Milan} (same as Google Cloud N2d) has 224 AMD Milan SMTs and 1.83 GB/s memory bandwidth per SMT. \textit{Skylake} (same as Google Cloud N1) has 112 Intel Skylake SMTs (2 sockets of 28 cores), 2.3 GB/s memory bandwidth per SMT.

Figure~\ref{fig:voxel-cpu-eff} shows the per-core performance when all cores independently run identical TPC-H query executions concurrently. For reference, we also measured the query execution performance when only one core is busy. When we benchmark systems with a single thread, the performance of AMD Milan and Intel Skylake is higher than that of the Smart NIC. 
When all cores run independent TPC-H query executions concurrently, the per-core performance of Intel IPU E2000 drops by 8--26\% (16 cores total). On the other hand, the per-core performance of x86 systems drops by 39\%--88\%. Across all cores on each system, AMD Milan shows 1.9-9.2x (median 4.7x) performance of E2000, and Skylake is 2.1-4.5x (median 3.6x) that of E2000. This suggests that a \sysname cluster with a $\phi$ of 3.6-4.7 might suffice to match the CPU performance of traditional servers.

The lone exception, the TPC-H Q6 query, performs a compute-bound scan of data in memory. The performance of Milan and Skylake drops mostly due to SMT core sharing.

\subsection{Higher End-Host Network Bandwidth}
\label{sec:network-bandwidth}

Relative to a traditional cluster, an important advantage of a \sysname cluster with $\phi > 1$ is the higher aggregate \blue{end-host network bandwidth} due to more Smart NICs. For big data workloads that involve large network transfers, a \sysname cluster can be cheaper and more energy-efficient: it can speed up network transmission to compensate for computation slowdown as a result of lower aggregate compute power. 

\begin{figure}
\centering
\includegraphics[width=\linewidth]{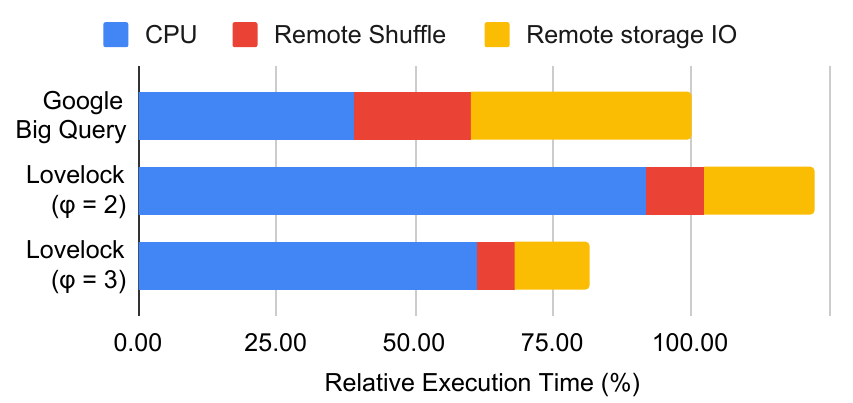}
\caption{Prediction of Big Query execution time with \sysname , based on the profiling data in~\protect\cite{hyperscaleBigDataProcessing}.
}
\label{fig:bqExecTime}
\end{figure}

A recently published breakdown of Google's BigQuery processing time~\cite{hyperscaleBigDataProcessing} reports that, on average, over 60\% of total time is spent on network operations, mainly remote shuffle and disaggregated storage IO. Using this breakdown, \sysname with $\phi > 1$ will provide higher network bandwidth, potentially reducing the remote shuffle and IO time. 

Figure~\ref{fig:bqExecTime} projects changes in BigQuery processing time with \sysname. To project CPU time, we multiply by 4.7, the median value of Milan's whole system CPU performance relative to E2000 in Figure~\ref{fig:voxel-cpu-eff}; \blue{then, we divide by $\phi$ since we assume linear speedup.} For remote shuffle and storage I/O time, we assume they are bottlenecked by network bandwidth. This is reasonable since BigQuery jobs usually scan terabytes or more of data, so shuffle and I/O involve large data transfers. Following~\cite{hyperscaleBigDataProcessing}, we attribute RPC processing at BigQuery workers to CPU times, not network transfers.

The first row in Figure~\ref{fig:bqExecTime} corresponds to the execution time composition reported in~\cite{hyperscaleBigDataProcessing}. We present two \sysname configurations: 2x and 3x more NICs than traditional servers (i.e., $\phi$ is 2 or 3). With $\phi=2$, total execution time increases by 22\% ($\mu=1.22$) \blue{because network overhead reduction is not enough to fully compensate $\frac{4.7 }{2} = 2.35$x reduction on aggregate CPU performance.} With $\phi=3$, total execution time will reduce by 19\% (i.e., $\mu=0.81$). 

For these two configurations, our model, together with cost and power values from~\cite{DpuWhitePaper}, suggests that \sysname's device cost advantage is 3.5x (respectively 2.33x) for $\phi$ of 2 (respectively 3). The 
energy savings are 4.58x for both.


\blue{Our model (\S\ref{sec:model}) ignores networking cost (fabric and ToR) increases for supporting more NICs. If we assume that networking accounts for 10\% of traditional cluster, Eq.~\ref{eq:cost} can be extended to $\frac{c_s + c_f + c_p}{\phi \cdot (1 + c_f) + c_p}$, where $c_f$ is the networking cost and may be assumed to be $c_s \times 10\% = 0.7$. With this updated cost model, the cost benefits with $\phi=2$ and $\phi=3$ will be 2.26x and 1.51x, respectively. We discuss this more in \S\ref{sec:discuss}.}

\blue{However, this analysis is pessimistic since it assumes fabric costs scale linearly with $\phi$. Instead, fabric capacity needs to increase only to keep up with execution time as determined by the slower CPUs. Thus, with $\phi=2$, the application slows down by $\mu=1.22$, so the fabric can actually be slower by about 19\% ($1-\frac{100\%}{122\%}$). Similarly, for $\phi=3$, the fabric needs to be faster by about 23\% to sustain the performance speedup. Thus, to sustain $\phi > 1$, it may not be necessary to provision $\phi$ times more capacity; rather it may be sufficient to over-subscribe the network.}





\subsection{Ability to Drive Accelerators}
\label{sec:accel-feasibility}

\sysname can benefit accelerator-based workloads in which (a) the CPU coordinates accelerator execution and data movement, and (b) accelerators are network-bound.

\vspace*{0.25ex}\noindent\textbf{CPU as coordinator}. In large language model training, CPUs effectively only coordinate training. To demonstrate that \sysname can lower the cost of this training, we trained large language models on 8 hosts each of which has 4 ML accelerators that can individually deliver about 50 TFLOPs. We used multiple model sizes, ranging from 1B to 39B, based on the configuration of dense models used in GLaM~\cite{glam}. The model parameters were evenly partitioned across the accelerators, and we set a global batch size of 64. With this training setting, we measured the resource usage of the hosts for 1,000 training steps. The role of CPU in this workload ranges from dispatching tasks to accelerators, checkpoining, and moving data across the network. The workload uses both inter-accelerator interconnect and datacenter network.

Table~\ref{tbl:tpuHostUse} shows the CPU and memory usage in host machines. Even the peak CPU use is well below the capacity of a smart NIC, IPU E2000. On average, training consumes only 3-5 GBs of memory, well below the capacity of a smart NIC. However, peak memory consumption can go up to twice the model size, when checkpointing the current training snapshots, including model parameters and optimizer states. We believe it is possible to reduce this peak by splitting model parameters into chunks and checkpointing a stream of these chunks. With this change, since an IPU E2000's DRAM capacity can be configured up to 48 GBs, each E2000 can drive 2-4 accelerators depending on the model size. 

Thus, \sysname with $\phi=1$ can likely sustain LLM training without any performance slowdown. 
Assuming that the device and energy cost of a host is 25\% of the entire system -- based on current servers with 4 GPUs -- and using cost and power values from~\cite{DpuWhitePaper} ($c_s$ = 7, $p_s$ = 11.2, $c_p$ = 21, and $p_p$ = 33.2), \sysname's cost advantage is 1.27x, and energy savings is 1.30x.

\begin{table}[t]
\scriptsize
\centering
\begin{tabular}{|l|r|r|r|r|r|}
\hline
Model & \makecell{Mean\\ CPU\%} & \makecell{Peak\\ CPU\%} & \makecell{Model Size\\ (per accel/Host)} & \makecell{Mean \\Mem Use} & \makecell{Max\\ Mem Use} \\ \hline
GLaM1B & 4.8\% & 8.9\% & 0.2GB / 0.8GB & 3.4GB & 5.0GB \\ \hline
GLaM4B & 3.8\% & 6.2\% & 0.4GB / 1.8GB & 3.8GB & 6.5GB \\ \hline
GLaM17B & 3.4\% & 10.2\% & 2.0GB / 8.1GB & 4.2GB & 17.8GB \\ \hline
GLaM39B & 2.1\% & 13.3\% & 4.5GB / 18.2GB & 4.7GB & 35.7GB \\ \hline
\end{tabular}
\caption{Host CPU and DRAM use during distributed training. ``CPU\%'' is normalized to the IPU E2000's CPU performance. CPU and memory use are sampled every minute from all 8 hosts, and avg and peak are calculated from the sampled data over the whole training.}
\label{tbl:tpuHostUse}
\end{table}

\vspace{0.25ex}\noindent\textbf{Higher aggregate network bandwidth.} Graph Neural Network (GNN) training is network bandwidth intensive. GNNs generate node embeddings from graph-structured datasets~\cite{gcn,graphsage,gat,clustergcn}. GNN computation requires significant network communication to preserve data dependencies in graphs~\cite{p3,pagraph,bgl}. For example, recent work~\cite{bgl} shows that creating one mini-batch requires fetching 200MB data from remote machines. While 8 V100 GPUs in one machine can compute 400 mini-batches per second, the shared 100Gbps network only allows 60 mini-batches, resulting in accelerator stalls and under-utilizing accelerators. 



Such stalls can also occur more generally in synchronous data-parallel training and model-parallel training/inference. Even if network bandwidth is provisioned enough to support average throughput, accelerators can still stall waiting for network transfers to complete. Such network stalls often account for over 20\% of execution time~\cite{pipedream, deeppool}, so providing 2x of bandwidth can easily bring 10\% speedup. A \sysname cluster with $\phi=2$, assuming accelerators account for 75\% of system power and cost (\S\ref{sec:model}), will have 1.22x cost and 1.4x power advantage over a traditional cluster.

\section{Discussion and Future Work}
\label{sec:discuss}

\parab{Improving Smart NICs for \sysname.}
Some smart NICs have limited memory bandwidth because their CPUs were designed to handle only subset of workloads. For example, Bluefield v3 has a memory bandwidth that is only 1.8x of network bandwidth (Table~\ref{tbl:bandwidthPerCompute}), so the internal CPU cannot process the data at line rate (IPU E2000 doesn't exhibit this limitation). Future NICs for \sysname can either allocate higher memory bandwidth or support DMA to PCIe devices.

\sysname can support extremely low latency networking because it does not incur a PCIe bus crossing between NIC and CPU (a special fabric is used instead). Current smart NIC hardware and drivers don't take advantage of this enough but can do so by directly writing to the internal CPU's cache line~\cite{nebula} or registers~\cite{nanopu}.

Memory on current smart NICs cannot support data-intensive workloads that rely on host memory for caching. We expect this limitation will disappear with CXL memory expansion (for in-memory caching) and swapping to far memory (for absorbing occasional surges).


\parab{Better isolation and performance predictability.} Because they have less-capable CPUs, a \sysname cluster can be efficiently utilized by a single application or a few applications of a single tenant.
This setting eliminates cross-tenant interference and improves performance predictability.
Avoiding host-level multi-tenant sharing also reduces the vulnerability of side channel attacks, thereby improving security isolation.

\parab{Scaling networking bandwidth.} One of the main benefits of \sysname is higher aggregate network bandwidth in configurations with $\phi > 1$. This can speed up applications (\S\ref{sec:network-bandwidth}) but not those that can exploit fast intra-host communication to reduce inter-host traffic. 
Consider the all-reduce step in ML training. In a traditional cluster, all GPUs within a host reduce gradients over fast inter-GPU interconnect (e.g., NVLink) before reducing across hosts over slow datacenter network. If a \sysname cluster scales by hosting fewer GPUs per smartNIC, the total datacenter network traffic for all-reduce operations will increase by $\phi$. 

Other data-intensive applications do not exhibit this behavior. Many data-intensive applications (e.g., Spark, BigQuery) are designed to use small-size worker nodes (4-16 cores per node/VM), and the number of worker nodes does not change, and neither would total network traffic, if these applications were hosted on a \sysname cluster. 

\parab{Scaling memory consumption.} In \sysname clusters with $\phi > 1$, the total memory consumed by application code and kernel will be higher than in a traditional cluster. We anticipate CXL-based memory disaggregation will alleviate this. A preliminary analysis of storage nodes shows that kernel's consumption is relatively small (1-2 GBs). We expect that memory used by applications will generally scale well since the input dataset is distributed, but this needs to be verified with additional analyses.

\parab{Networking and RPC performance.} Smart NICs ASICs can offload packet processing and congestion control (\S\ref{sec:smartnic}), but likely not other networking components or RPC services. These may need to run on smart NIC CPUs, and we believe networking and RPC services can be optimized to run on these CPUs. For example, eRPC~\cite{erpc} demonstrates that a single core can achieve 10 million small RPCs per second or 75 Gbps with large messages. Our preliminary experiments with IPU E2000 suggest that a single ARM core can sustain over 25 Gbps with large message RPCs.

\parab{Data processing accelerators.}  Beyond network acceleration, smart NICs also have fixed-function data processing accelerators for crypto, compression, CRC, and copy. These can support infrastructure services like full-featured RPC, data center file systems, and logging without significantly taxing the smart NIC CPU and DRAM resources.

\parab{Network cost modeling.} \blue{We can extend our model (\S\ref{sec:model}) to reflect the increased cost of the network fabric when $\phi > 1$ by assuming that network cost scales linearly with cluster size (\S\ref{sec:network-bandwidth}). However, this is pessimistic; smaller capacity increases might suffice to sustain application speedups, since applications will be slowed down by the smaller CPU. In addition, rack-local disaggregation can further reduce additional fabric capacity required. Right-sizing fabrics for applications will be crucial for \sysname viability.}

\section{Related work}

\parab{Offloading to smart NICs from host.} Smart NICs are designed to offload various networking functions from host cores, and their effectiveness over general cores is proven~\cite{accelnet, DpuVmwareRedis, DpuWhitePaper}. Other research has explored the potential of offloading user applications beyond networking functions~\cite{floem, ipipe}, and demonstrated potential energy savings~\cite{e3} and lower tail latency~\cite{argus}. But, to our knowledge, no prior work has proposed replacing servers with smart NICs, as \sysname does.

\parab{Disaggregated datacenter.} 
Rapid improvements in network bandwidth and latency has enabled resource disaggregation~\cite{netreq4disagg, quicksand, nu, aifm, carbink}. For efficient disaggregation, prior work explores custom hardware and software for non-compute nodes (e.g., memory node)~\cite{legoos, cilo, skadi, fullydisagg}. Relative to \sysname, these hardware-based disaggregation approaches require enormous software restructuring both in user applications and infrastructure applications. 
\clearpage
\bibliographystyle{abbrv} 
\begin{small}
\bibliography{main}
\end{small}

\end{document}